\documentclass[11pt]{article}

\usepackage{epsfig}

\begin{document}
\begin{titlepage}
\begin{center}
{\Huge Recent Lattice QCD Results from the UKQCD Collaboration}

\vspace{8mm}
{\Large Chris Allton}

\vspace{8mm}
Department of Physics, University of Wales, Singleton Park, Swansea
SA2~8PP, UK\\
\vspace{8mm}
c.allton@swan.ac.uk
\vspace{8mm}
\end{center}

\begin{abstract}
The lattice technique of studying the strong interaction of matter is
used to obtain predictions of the hadronic spectrum. These simulations
were performed by the UKQCD collaboration using full (unquenched) QCD.
Details of the results, a comparison with quenched data, and novel
methods of extracting spectral properties are described.
\end{abstract}

\end{titlepage}


\section{Introduction}

The lattice technique of studying field theories such as QCD (the
strong force of particle physics) has had a long and successful
history\cite{lat99}. In this method, the usual computational modelling
trick of discretising space-time, and replacing derivatives by finite
differences is employed. This allows the study of field theories at
any value of the coupling, and contrasts with perturbation theory
which is restricted to systems where the interaction strength is weak.
For this reason, the lattice technique is particularly applicable to
calculations involving the strong interaction of particle physics, and
is now regarded as the method of choice for its study. It has the
advantage over of its competitors in that it has no model assumptions,
and is also systematically improvable.

Historically, due to computational resource issues, lattice simulations
have used the ``quenched'' approximation (which ignores virtual
quark loops in the vacuum). Now that computer power is in the
Teraflop region, lattice QCD studies are able to simulate without this
approximation.

The UKQCD collaboration\cite{ukqcdweb} has played a leading role in the
study of QCD via the lattice technique. Most of the
present emphasis of this collaboration is the study of hadrons using
``full'' (i.e. ``unquenched'') QCD. A summary of recent results will
be given (see \cite{ukqcd1,ukqcd2} for details) and an exciting new
approach to obtaining spectral information from the lattice based on
the ``Maximum Entropy Method'' will be discussed.

The main computer resource of the UKQCD collaboration is currently a
Cray T3E in Edinburgh, however this will soon be strengthened by the
acquisition of an APEmille computer in Swansea. Furthermore, funding
has been approved for a supercomputer in the multi-Teraflop range in
the medium term. 


\section{Brief Overview of Lattice Gauge Theory}
\label{sec:lgt}

In lattice QCD, the quark degrees of freedom are defined on the sites of the
4-dimensional lattice, and the gluonic degrees of freedom (corresponding to
the carriers of the strong force) are defined on the links. This enables a
lattice version of the theory to be defined which maintains as many
of the symmetries of the continuum theory as possible. (For a reviews of
lattice QCD see \cite{lgt_reviews}.)

Typical quantities of interest can then be expressed as $n$-point
functions, $G_n(t)$, of hadronic operators. Specifically these are
weighted averages over configurations of the degrees of freedom with a
Boltzmann-like weight, $e^{-\beta S}$ where $S$ is the lattice version
of the continuum QCD action. It is important to note that this is
exactly analogous to statistical mechanics. This means that many of the
techniques developed in that area are applied to lattice QCD -- for
example the use of the Monte Carlo integration method.

Quantities of physical interest, such as hadronic masses, can be extracted from
calculations of the 2-point function $G_2(t)$ as follows. It can be shown that
\begin{equation}\label{eq:decompo}
G_2(t) = \sum_i Z_i e^{-M_i t}
       \rightarrow Z_0 e^{-M_0 t} \;\;\; \mbox{as } t\rightarrow \infty.
\end{equation}
The sum is over all states $|i>$ with the same quantum numbers as the
operators which define the 2-point function, $G_2(t)$.
Hence the ground state mass of a hadron, $M_0$, can be extracted simply
by fitting $G_2(t)$ to an exponential at sufficiently large times $t$.

This is not the end of the story though because $M_0$ (and any other
quantity extracted from a lattice calculation) is a function of the
input parameters of the simulation.  These are the coupling strength
$g_0$, the quark mass $m$ and the volume $L^3$.  Ideally what is
required is the continuum limit ($a \rightarrow 0$) of $M_0$ which
corresponds to $g_0 \rightarrow 0$ (since QCD is asymptotically
free). However, as has been found by several groups
(e.g. \cite{sesam,ukqcd1}), the quark mass $m$ also has impact on the
lattice spacing $a$, i.e. $a = f(g_0,m)$. This will have an important
implication as we will see in a moment.

Another issue regarding lattice QCD is the error which is introduced in
the discretisation of the continuum theory. In the conventional approach, the
lattice action reproduces the continuum action only up to terms of
${\cal O}(a)$. Over the last several years, lattice groups, including
UKQCD, have been using so-called {\em improved} versions of the
lattice action which have lattice errors of ${\cal O}(a^2)$.
While this is an obvious improvement over conventional lattice
actions, it does not preclude lattice results being contaminated
with lattice artefacts, all-be-it at the ${\cal O}(a^2)$ level.

In order to establish a clean signal for many studies it is imperative
to fix $a = $ {\em const} for two reasons: (i) as seen above, the
lattice calculation leaves the answer correct to ${\cal O}(a^2)$, so in
order to keep this error fixed, $a$ must obviously be fixed; (ii) in
order to maintain fixed finite-size effects, $a$ must be fixed (simply
because the lattice size $L = Na$ where $N$ is the number of lattice
sites along a side). Thus, considering that $a = f(g_0,m)$ (see above)
the requirement that $a = $ {\em const} defines a trajectory in the
$(g_0,m)$ plane.



\section{Results}

In the previous section we argued that simulating along a fixed $a$
trajectory in $(g_0,m)$ parameter space lead to the cleanist
study. UKQCD have followed this philosophy and performed simulations
at four different parameter choices for $(g_0,m)$ such that $a = $
{\em const} $\approx 0.1 \;fm$ (see \cite{ukqcd2} for details).
Note that one of these four parameter choices corresponds to the
quenched approximation (specifically $m = \infty$).
It is important to note that when a physical quantity changes value
between these four parameter values, the cause can be attributed to
variations in the quark mass $m$ (i.e. ``unquenching'' effects), and
{\em not} lattice systematics such as finite volume effects (see
previous section).

In this section an overview of three quantities of particular interest
is given.


\subsection{Static Quark Potential}

One of the fundamental quantities that can be studied on the lattice
is $V(r)$, the potential between two infinitely heavy quarks (the ``static
quark potential''). Phenomenologically,
\begin{equation}
V(r) = V_0 + \frac{\alpha}{r} + \sigma r.
\label{eq:statpot}
\end{equation}
In Figure 1, $V(r)$ is plotted for all four parameter choices. The
phenomenological model Eq.(\ref{eq:statpot}) is also shown. As can be
seen there is agreement amongst the data and Eq.(\ref{eq:statpot}) to a
high precision. However, close inspection of the data near the
origin shows a systematic tendency towards a higher value of $\alpha$
as the quark mass $m$ is decreased towards it physical value. We take
this as a signal for unquenching effects \cite{ukqcd2}.




\begin{minipage}[h]{92mm}
\vspace{5mm}
\epsfig{file=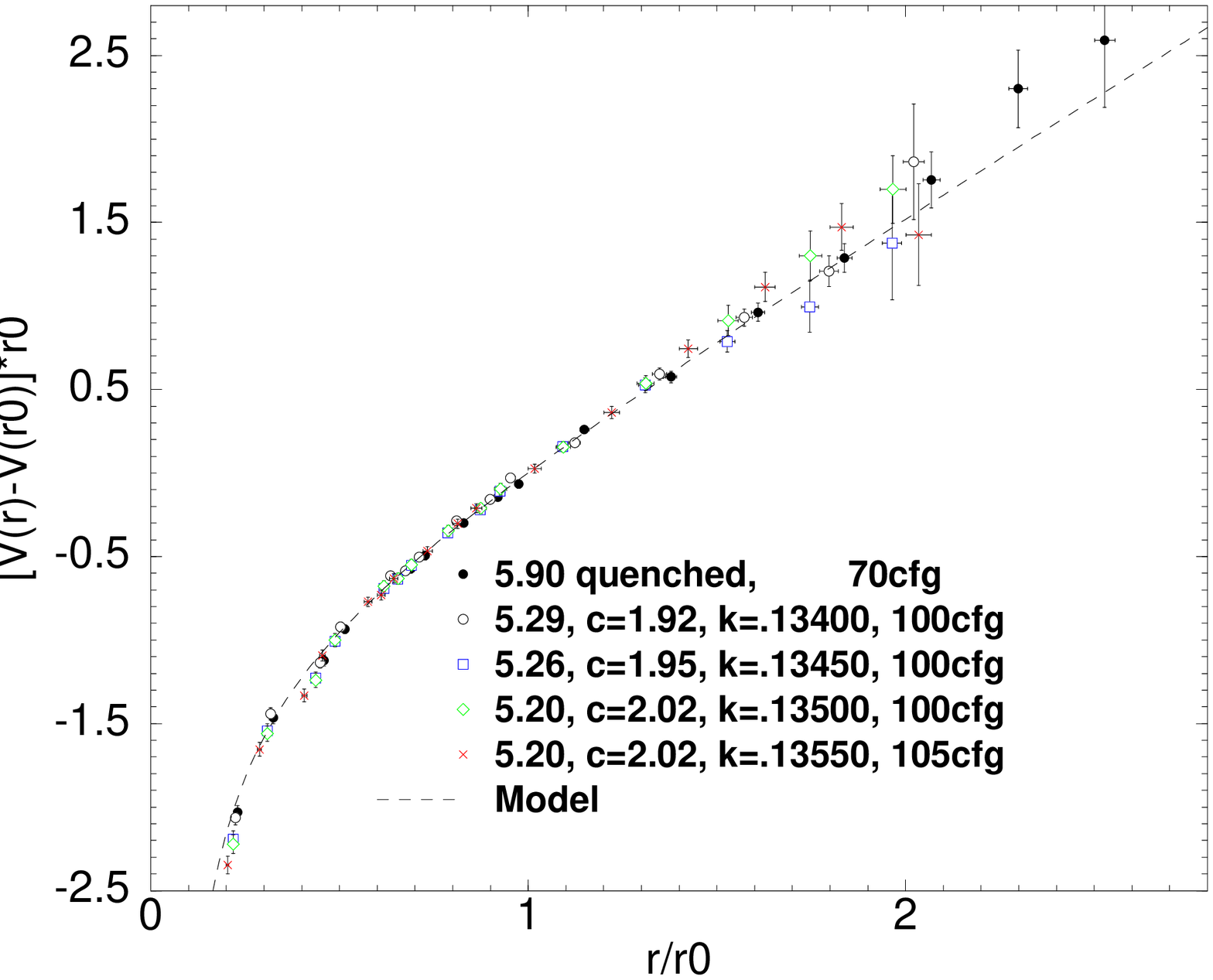,height=9cm}
\end{minipage}
\begin{minipage}[h]{92mm}
\vspace{-25mm}
\epsfig{file=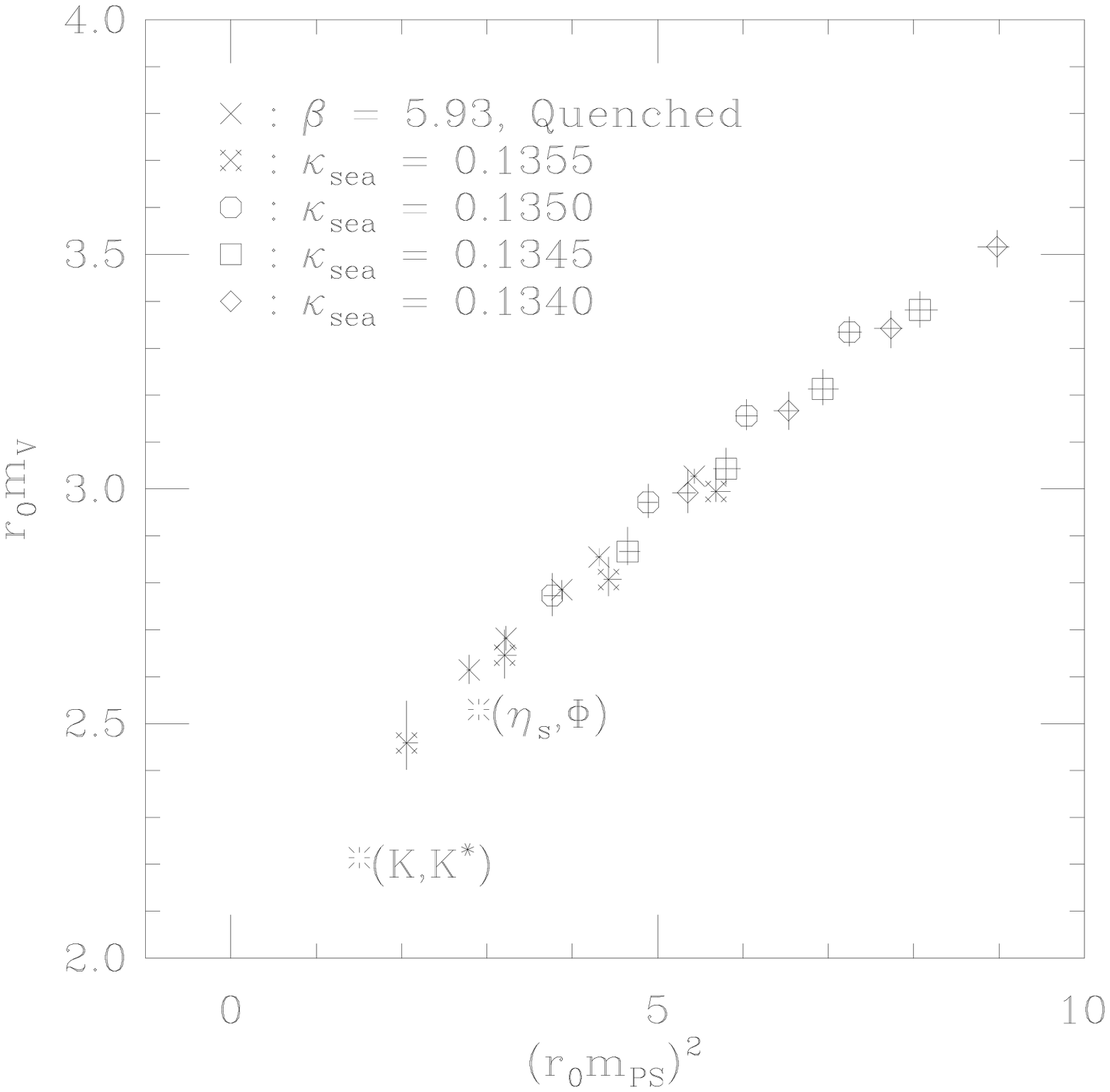,height=6.8cm}
\end{minipage}

\vspace{-25mm}
Figure 1: {\em Static quark potential versus distance $r$ on left,
and Hadronic masses $M_V$ versus $M_{PS}^2$ on right. All quantities
have been scaled by the distance scale $r_0$.}



\subsection{Hadronic Spectrum}

In this subsection the hadronic spectrum is discussed. We concentrate
here on the pseudoscalar and vector mesonic states whose masses are
denoted $M_V$ and $M_{PS}$ respectively, since these can
be determined most unambiguously.
Theoretically we expect the following relationship to hold:
\[
M_V = A + B \; M_{PS}^2,
\]
where $A$ and $B$ are constants. In Figure 1, $M_V$ is plotted against
$M_{PS}^2$ for all four parameter choices. Note that within each
parameter choice, the ``valence'' quark mass is still a free parameter
hence the multiple data points for each parameter choice in the figure
\cite{ukqcd1}. Also plotted in Figure 1 is the experimental values for
the $\pi, \rho, K, K^\ast, \eta_s$ and $\phi$ mesons \cite{pdb}. As can
be seen there is a trend towards the experimental values as the
quenched approximation is removed.




\subsection{Topological Susceptibility}

The topological susceptibility, $\chi$, plotted in Figure 2, shows
a marked unquenching effect. This is interpreted as a suppression of
instanton effects by the light quark modes \cite{top}. Of the three
quantities discussed in this section, $\chi$ shows by far the most
significant unquenching features.

\begin{figure}[htp]
\begin{center}
\epsfig{file=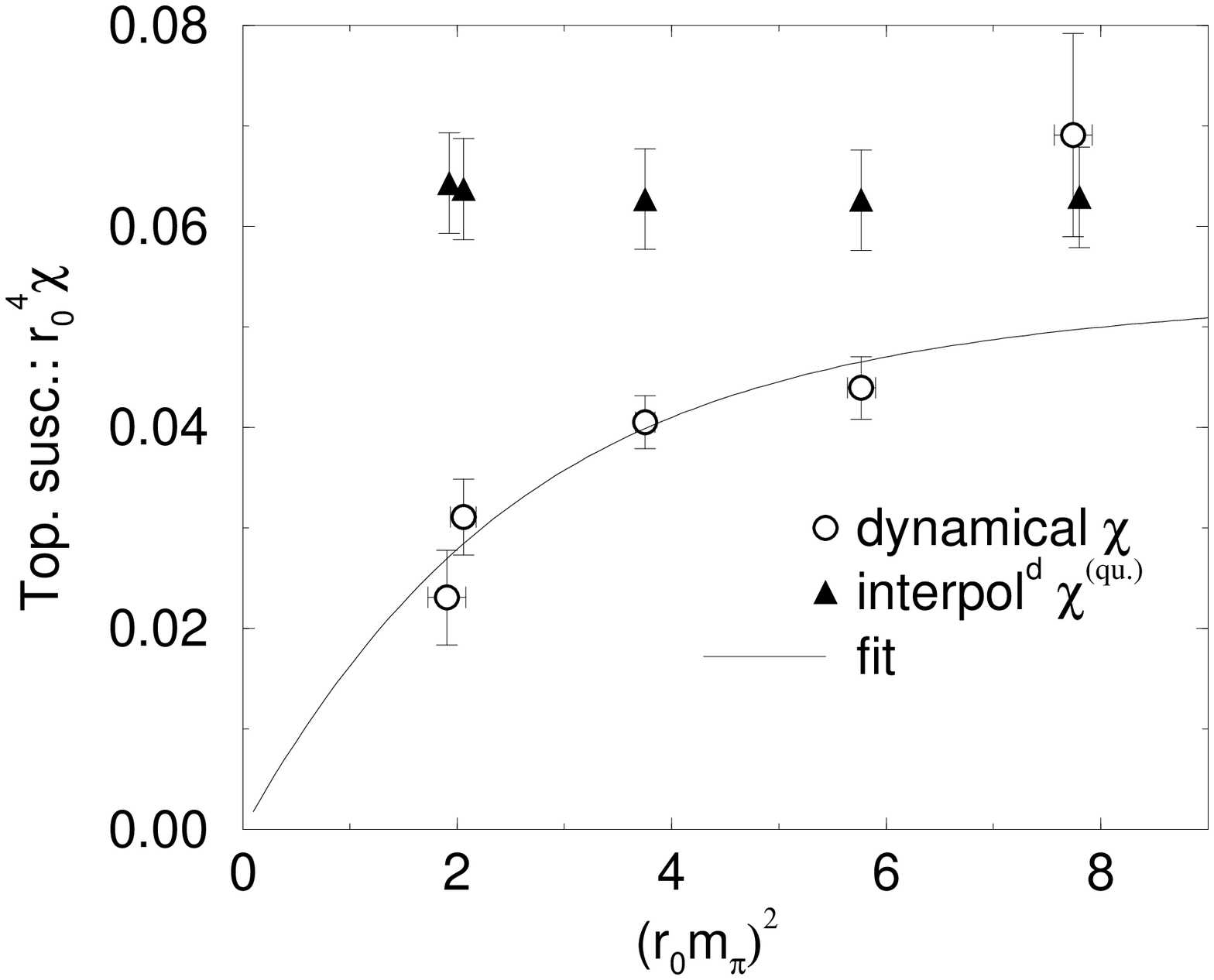,height=6cm}
\vspace{-5mm}
\protect\label{fig:top}
\end{center}
\vspace{1mm}
Figure 2: 
{\em Topological susceptibility, $\chi$, versus (effective) quark
mass. All quantities have been scaled by the distance scale $r_0$.}
\end{figure}



\section{Spectral Functions}

The 2-point function, $G_2(t)$ can be spectrally decomposed
\[
G_2(t) \equiv \int_0^\infty K(t,\omega) \rho(\omega) d\omega,
\]
where $K$ is the Kernel (it is $e^{-\omega t}$ in our case - see 
Eq(\ref{eq:decompo})) and $\rho(\omega)$ is the Spectral Function.
$\rho(\omega)$ contains information about the ground and excited
states and the continuum for the particular channel of interest.

In \cite{mem} the ``Maximum Entropy Method'' which is based on Bayesian
statistics was employed to extract $\rho(\omega)$ directly from the
lattice $G_2(t)$ data. This is an exciting development which we plan to
implement since a huge wealth of information becomes available once
$\rho(\omega)$ is calculated. This is especially true in QCD in the
high temperature phase where many channels no longer have bound states,
and the traditional approach (detailed in section 2) is simply not
applicable.



\section{Summary}

The UKQCD collaboration have made important progress in lattice QCD by
implementing a ``matched'' trajectory approach where the lattice
spacing is kept fixed, and thus the unquenching, finite-size and
discretisation effects are dis-entangled. A summary is given here of
the results for the static quark potential, hadronic spectrum and
topological susceptibility - see \cite{ukqcd1,ukqcd2} for full
details. A new and very promising method which enables the direct
extraction of the spectral function from lattice data is discussed.





\begin{thebibliography}{9}

\bibitem{lat99}
see e.g. {\em Lattice'99, Proceedings of the XVIIth International
Symposium on Lattice Field Theory,} Pisa, Italy, 29 June - 3 July 1999.
Nucl. Phys. {\bf B (Proc. Suppl.) 83-84} (April 2000), and the
electronic archive site: {\tt http://xxx.lanl.gov/archive/hep-lat}.

\bibitem{ukqcdweb}
{\tt http://www.ph.ed.ac.uk/ukqcd/index.html}

\bibitem{ukqcd1}
UKQCD Collaboration (C.R. Allton et al.),
{\em Light hadron spectroscopy with O(a) improved dynamical fermions},
Phys.\ Rev.\ {\bf D60} (1999) 034507 {\tt hep-lat/9808016}

\bibitem{ukqcd2}
UKQCD Collaboration (C.R. Allton et al.),
{\em Effects of non-perturbatively improved dynamical fermions in lattice QCD},
in preparation.

\bibitem{lgt_reviews} see e.g.
Stephen R. Sharpe, Plenary talk at ICHEP98, {\tt hep-lat/9811006};
Rajan Gupta, {\em Parallel Computing} (special issue devoted to Lattice QCD),
{\tt hep-lat/9905027};
V. Lubicz, Invited talk at the XX Physics in Collision Conference, June 29 - July 1 2000, Lisbon, Portugal,
{\tt hep-ph/0010171}.

\bibitem{sesam} SESAM-Collaboration, N.Eicker et al,
Phys.Lett. B407 (1997) 290.

\bibitem{pdb}  Review of Particle Physics (Particle Data Book),
Eur. Phys. J. {\bf C 15} (2000) 1.

\bibitem{top} UKQCD, A. Hard and M. Teper hep-ph/0004180.

\bibitem{mem} Y. Nakahara, M. Asakawa and T. Hatsuda,
Phys. Rev. {\bf D60} (1999) 091503,
Nucl. Phys. (Proc.Suppl.) {\bf 83} (2000) 191.

\end{thebibliography}
\end{document}